\documentclass[a4paper,11pt]{article}
\pdfoutput=1 

\usepackage{jcappub} 
\usepackage[T1]{fontenc}
\usepackage{bm}
\usepackage{makecell}
\usepackage{multirow}

\newcommand{\LCDM}{$\Lambda$CDM}
\newcommand{\fLCDM}{$f\Lambda$CDM}
\newcommand{\gLCDM}{$\gamma\Lambda$CDM}
\newcommand{\mpc}{\mathrm{Mpc}}
\newcommand{\gpc}{\mathrm{Gpc}}
\newcommand{\op}{\mathcal{O}}
\newcommand{\linear}{\mathrm{L}}

\newcommand{\prob}{\mathcal{P}}
\newcommand{\tracer}{\mathrm{g}}
\newcommand{\tracerdet}{\mathrm{g,det}}
\newcommand{\tracerdensityRSD}{\tilde{\delta}_\tracer}
\newcommand{\matter}{\mathrm{m}}
\newcommand{\clusteringamplitude}{A_\mathrm{s}}
\newcommand{\noiseamplitude}{\sigma_\epsilon}
\newcommand{\fiducial}{\mathrm{fid.}}
\newcommand{\LEFTfield}{\texttt{LEFTfield}}
\newcommand{\CLASS}{\texttt{CLASS}}
\renewcommand{\vec}[1]{\bm{#1}}
\newcommand{\explicit}{\mathrm{e}}
\newcommand{\marginalized}{\mathrm{m}}
\newcommand{\pnameAkitsu}{Akitsu et al.}
\newcommand{\pnameNguyen}{Nguyen et al.}
\newcommand{\pnameBeyond}{Beyond-2pt Challenge}

\title{Cosmological information from field-level analysis in redshift space}

\author[]{Julia Stadler}
\affiliation[]{University Observatory, Faculty of Physics, Ludwig-Maximilians-Universität, Scheinerstr. 1, 81679 Munich, Germany}
\affiliation[]{Max-Planck-Institut für Astrophysik, Karl-Schwarzschild-Str. 1, 85748 Garching, Germany}
\affiliation[]{Excellence Cluster ORIGINS, Boltzmannstr. 2, 85748 Garching, Germany}
\emailAdd{j.stadler@lmu.de}

\abstract{
The field-level analysis of galaxy clustering with a forward model based on the Effective Field Theory of Large Scale structure provides cosmological measurements at optimal statistical precision which are simultaneously robust with respect to theoretical uncertainties regarding galaxy formation. Yet, the quantitative gains of field-level analysis in the perturbative regime over the current state-of-the-art, set by the combination of power- and bispectrum, are not obvious, and they depend on analysis choices, the cosmological model, and the observed galaxy population. We therefore quantify the precision improvement from field-level analysis over power- and bispectrum from redshift-space snapshots that closely resemble the DESI LRG sample. Our mock data are created with the \texttt{LEFTfield} forward model, which we also use for the subsequent field-level analysis and for a Fisher forecast of the power- and bispectrum precision. We find significant improvements from the field-level analysis over the power- and bispectrum for a \LCDM{} cosmology; in particular the primordial fluctuation amplitude can be measured twice as precisely. Precision gains at the field-level are even more pronounced for extended cosmologies with a more flexible dependence between power spectrum shape and structure growth; they reach up to a factor of three in the primordial amplitude and a factor of two in parameters that capture deviations from \LCDM{}. Our results highlight the potential of field-level analysis for future discoveries. Realizing this potential will require substantial developments on the modeling of observations and observational systematics as well as on the numerical inference techniques.
}

\begin{document}
\maketitle
\flushbottom

\section{Introduction}

An ambitious program of stage IV surveys \cite{DESI:2016fyo, Amendola:2016saw, 2014PASJ...66R...1T, Dore:2014cca, LSST:2008ijt, 2019arXiv190205569A} aims at unveiling the nature of Dark Energy by measuring its impact on the expansion of the Universe and the growth of structure therein. The analysis of these data poses a double challenge: how can information be extracted optimally while drawing conclusions that are robust to modeling uncertainties. Put more profane, how to minimize the statistical uncertainty without underestimating the systematic error. In galaxy clustering, where observed galaxies are used as tracers of the matter distribution, a major source of theoretical uncertainty is the bias relation.

The Effective Field Theory of Large Scale Structure (EFT of LSS) \cite{Baumann:2010tm, Carrasco:2012cv, Hertzberg:2012qn, Carroll:2013oxa, Senatore:2014eva, Goldberger:2004jt, Mirbabayi:2014zca} is the most general description of the matter-galaxy connection on large scales (see \cite{Desjacques:2016bnm} for a review). The galaxy density $\delta_\tracer$ is expressed as a functional in local gravitational observables (or bias operators) $\op$ that are computed from the matter density $\delta_\matter$. Expanding the functional perturbatively,
\begin{equation}
\delta_\tracer \left[\delta_\matter\right] = \sum_\op\, b_\op \op\left[\delta_\matter\right] + \epsilon_\tracer \,,
\label{eq:intro-bias}
\end{equation} 
yields a series that can be truncated at finite order. The bias coefficients $b_\op$ are free parameters. The galaxy density is split into a deterministic part, $\delta_\tracerdet = \sum_\op b_\op\op$, and a stochastic part, $\epsilon_\tracer$. The stochastic part can itself be split into operators, and the leading contribution is a white-noise field \cite{Desjacques:2016bnm, Cabass:2019lqx, Rubira:2024tea, Rubira:2025rqo}. Because the expansion contains all possible terms permitted by physical symmetries, it is the most general ansatz under the assumption that gravity is described by General Relativity. Thus, the EFT bias expansion is very robust to model misspecification. However, the bias expansion can only be applied on scales where the perturbative treatment is justified ($k \ll k_\mathrm{NL} \simeq 0.25\,h/\mpc$ at $z=0$).

The perturbative description of galaxy bias naturally motivates to analyze observations in terms of $n$-point correlation functions. Structure formation proceeds bottom-up, and on very large scales the leading linear term  $\delta_\tracerdet \simeq b_1 \delta_\linear$ together with Gaussian noise sufficiently captures the galaxy density. Here, $\delta_\linear$ denotes the linear matter density, obtained when extrapolating primordial perturbations to the present epoch with linear growth. It follows a Gaussian distribution very closely \cite{Planck:2019kim}, such that the power spectrum captures all information on the cosmological parameters. The higher-order bias terms in eq.~\eqref{eq:intro-bias} and higher-order contributions to the gravitational evolution of $\delta_\matter$ carry additional, non-Gaussian information that can be mapped to successively higher-order $n$-point functions \cite{Cabass:2023nyo, Schmidt:2025iwa}. It has also been argued that the signal-to-noise ratio at which $n$-point functions can be measured is increasingly suppressed with higher $n$ \cite{Spezzati:2025zsb}. The present state-of-the art in the analysis of observations is the combination of power- and bispectrum (P+B). The power spectrum analysis has been successfully validated in a parameter-masked mock challenge \cite{Nishimichi:2020tvu}, applied to SDSS/BOSS data \cite{Ivanov:2019pdj, DAmico:2019fhj} and are now the standard approach to the full-shape analysis in DESI \cite{Maus:2024sbb, DESI:2024hhd, DESI:2024jxi}. The bispectrum was e.g. validated in \cite{Beyond-2pt:2024mqz}, applied to BOSS \cite{Philcox:2021kcw, DAmico:2022osl}, and to DESI \cite{Chudaykin:2025aux, Novell-Masot:2026sgg, Chudaykin:2026nls, Ivanov:2026dvl, Chudaykin:2025lww, Verdiani:2025jgz}.

An alternative analysis strategy is to forgo the compression into $n$-point functions entirely, and to consider the three-dimensional (coarse grained) galaxy density field as data vector. In this case, the primordial perturbations become additional free parameters and are inferred jointly with bias coefficients and cosmological parameters. Such field-level analysis \cite{Jasche:2012kq, Elsner:2019rql, Schmidt:2018bkr, Schmidt:2020viy, Ramanah:2018eed} can access all information (up to a given cut-off scale) and guarantees optimal constraints. The central link between data and theoretical model prediction is the field-level likelihood, which follows from the stochastic part of the galaxy bias expansion. At leading order in the noise, it is \cite{Cabass:2019lqx, Schmidt:2020tao}
\begin{equation}
-\ln \prob\left(\delta_\tracer | \delta_\tracerdet\left[\delta_\linear, \theta, b_\op \right] \right) = \frac{1}{2} \sum_{\vec{k}\neq 0}^{k_\mathrm{max}} \left[ \ln\left(2\pi\noiseamplitude^2\right) + \frac{\left|\delta_\tracer\left(\vec{k}\right) - \delta_\tracerdet\left(\vec{k}\right)\right|^2}{\noiseamplitude^2}\right] \,,
\label{eq:intro-likelihood}
\end{equation}
where the noise amplitude $\noiseamplitude$ is a free parameter similar to the bias coefficients. The upper limit to the sum, $k_\mathrm{max}$, implements the cut-off scale and guarantees that only modes under perturbative control are included. It also sets the (minimum) resolution at which the galaxy density has to be represented. 

Formally, cosmological parameters constraints follow from marginalizing the joint posterior over initial conditions and bias coefficients,
\begin{equation}
\prob\left(	\theta | \delta_\tracer \right) \propto \int db_\op ~ \int \mathcal{D}\delta_\linear ~ \prob\left(\delta_\tracer | \delta_\tracerdet\left[\delta_\linear, \theta, b_\op \right] \right) \, \prob\left(\delta_\linear | \theta\right) \, \prob\left(\theta, b_\op\right).
\label{eq:intro-cosmologyposterior}
\end{equation}
This posterior describes a hierarchical Bayesian model of cosmic large-scale structure, where the prior on the linear density field follows from the cosmological model. In practice, the marginalization over initial conditions in eq.~\eqref{eq:intro-cosmologyposterior} is performed numerically after sampling from the high-dimensional joint posterior. Field-level analysis therefore provides posterior samples of the initial conditions, the structure formation history, and matter distribution over the analyzed volume with consistent cosmology dependence as a by-product. They allow for cross-validation and follow-up studies (see e.g. \cite{Lavaux:2015tsa, Jasche:2014vpa, Lavaux:2019fjr, Andrews:2026okr, McAlpine:2025uzh, Nguyen:2020yuc, Bartlett:2022ztj, Kostic:2023arx} for some examples which use the initial conditions posterior at a fixed cosmology). The price to pay for optimal cosmological measurement and access to the initial conditions is a considerable computational overhead. Evaluation of the posterior requires to forward model $\delta_\tracerdet$, and during the inference those parameter combinations of $\theta$, $b_\op$ and $\delta_\linear$ are explored which give a good description of the data. In particular $\delta_\linear$, which is represented on a three-dimensional grid, can easily have millions of free parameters. Inference over such a high-dimensional space is only feasible with advanced sampling techniques such as Hamiltonian Monte Carlo (HMC, \cite{Neal:2011mrf}).

It is clear that cosmology constraints from field-level analysis are optimal, although the precise gains over power- and bispectrum are debated in the literature. While \pnameAkitsu{} \cite{Akitsu:2025boy} argues for moderate improvements, \pnameNguyen{} \cite{Nguyen:2024yth} and the \pnameBeyond{} \cite{Beyond-2pt:2024mqz} found more substantial gains. It is clear that the answer to \textit{``How much more information can be extracted from galaxy clustering at the field level?''} depends on the scenario considered, such as tracer density, volume, cut-off scale but also the cosmological model and which parameters are varied. The different conclusions between \pnameNguyen{}, the \pnameBeyond{}, and \pnameAkitsu{} have in parts been attributed to the different orders these works adopted for the bias expansion. Further, \pnameNguyen{} and the \pnameBeyond{} focus on field-level analysis in the tracer restframe and find the largest improvements for the clustering amplitude $\sigma_8$. In the restframe, field-level analysis is understood to effectively break the bias-amplitude degeneracy with higher-order information. In redshift-space, where galaxies actually are observed, the degeneracy is already lifted by anisotropies.

In this work, we compare the information from field-level analysis to the power- and bispectrum on a redshift-space mock that closely resembles the DESI LRG sample. All our model predictions are obtained with the \LEFTfield{} code \cite{Schmidt:2020ovm, Stadler:2024aff, Stadler:2024fui}. While previous works with \LEFTfield{} varied the clustering amplitude and the bias coefficients, we here include the full flat \LCDM{} parameter space and some extensions. The work is structured as follows; we discuss our cosmological model choices, the field-level analysis with \LEFTfield{}, and the methodology for the power-and bispectrum in section~\ref{sec:methods}. Section~\ref{sec:data} explains the mock data, and section~\ref{sec:results} shows and discusses our results. We conclude in section~\ref{sec:conclusions} and highlight future challenges. 

\section{Methods}
\label{sec:methods}

To investigate the information content of redshift-space snapshots, we compare field-level analysis to a Fisher forecast \cite{Tegmark:1997rp} of the power- and bispectrum. For both, we use the same cosmology, forward model, scale cuts and fiducial values.  The Fisher forecast generally yields a lower limit on the achievable precision (the Cramér-Rao bound), and for a linear model with an unbiased Gaussian likelihood it predicts the true precision \cite{Verde:2009tu}. In fact, the assumption of a Gaussian likelihood is very common to power- and bispectrum analyses. Moreover, the comparison between Fisher forecast and simulation-based inference \cite{2020PNAS..11730055C}, which avoids any assumptions on the form of the likelihood, revealed good agreement on restframe snapshots \cite{Tucci:2023bag}. Therefore, we expect an accurate estimate of the constraining power from the Fisher forecast. In the worst case, the Fisher forecast will give too much credit to power- and bispectrum over field-level analysis.

We discuss the details of our methodology in the following, starting with the cosmological model choice in subsection~\ref{sec:methods-cosmology}. The \LEFTfield{} forward model and the field-level analysis are summarized in subsection~\ref{sec:methods-fieldlevel}, and the power- and bispectrum forecast with \LEFTfield{} in subsection~\ref{sec:methods-fisher}.

\subsection{Cosmological model choice}
\label{sec:methods-cosmology}

The data considered are redshift-space snapshots of the galaxy density (see section~\ref{sec:data} for details). They contain information on the shape and amplitude of galaxy clustering and from anisotropies induced by redshift-space distortions. The snapshots lack geometric information from the projection of physical distances onto angles, most notably the angle under which Baryon Acoustic Oscillations (BAO) are observed, and from Alcock-Paczynski distortions more general \cite{Alcock:1979mp}. To explain our cosmological model choices, we discuss in the following which cosmological parameters can be meaningfully constrained by the available information (see also \cite{Ivanov:2019pdj, Beyond-2pt:2024mqz} for similar discussion). 

In a \LCDM{} cosmology, there are five cosmological parameters that can in principle impact cosmic large-scale structure,
\begin{equation}
\omega_\matter\,, \quad \omega_\mathrm{b}\,, \quad h\,, \quad n_\mathrm{s}\,, \quad \textrm{and} \quad \clusteringamplitude\,.
\label{eq:method-lcdmparameters}
\end{equation} 
The latter two, the scalar amplitude $\clusteringamplitude$ and the spectral index $n_\mathrm{s}$ define the amplitude and shape of the primordial curvature perturbations. The present day Hubble rate in units of $100 \times \mathrm{km} /\mpc/\mathrm{s}$ is denoted by $h$. Finally, $\omega_\mathrm{b}$ refers to the physical density of baryons at redshift zero, while $\omega_\matter$ is the physical matter density including baryons and Dark Matter. The energy density associated with the cosmological constant follows from demanding a Universe with zero curvature. 

In terms of parameter constraints, the shape of the linear power spectrum depends sensitively on $\omega_\matter$ and also on the baryon density and the spectral index. Nevertheless, constraints from galaxy clustering on $\omega_\mathrm{b}$ and $n_\mathrm{s}$ are rather weak. The amplitude of the galaxy power spectrum to linear order is proportional to the parameter combination $ b_1^2 \, \clusteringamplitude\, D^2(a) $, with the linear growth factor $D(a)$ given by
\begin{equation}
D(a) = \frac{5 \Omega_\matter}{2}\,\frac{H(a)}{H_0} \int_0^a da' ~ \frac{H_0^3}{\left(a' H(a')\right)^3} \,.
\label{eq:methods-grothfactor}
\end{equation}
In the galaxy rest frame, the degeneracy between $\clusteringamplitude D^2(a)$ and $b_1^2$ is broken by higher order information, while the degeneracy between primordial amplitude and growth persists at all perturbative orders. Redshift-space distortions contain information on the growth rate,
 \begin{equation}
f(a) = \frac{\partial \ln D(a)}{\partial \ln a} \,.
\label{eq:mathods-growthrate}
\end{equation}
They introduce an anisotropic clustering signal that can be probed by multipoles of the galaxy power spectrum. At the linear level, the monopole measures $b_1 \sigma_8$ and its ratio to the quadrupole $f \sigma_8$. Here, $\sigma_8$ denotes the root-mean-square of matter fluctuations on scales of $8\,\mpc/h$ which scales directly with $\clusteringamplitude D^2(a)$. Because $f$ is determined from $\Omega_\matter = \omega_\matter/h^2$, redshift-space distortions break the bias-amplitude degeneracy in \LCDM{}.

From the previous discussion, it is clear that redshift-space snapshots meaningfully constrain $\omega_\matter$ and $\clusteringamplitude$. We choose to fix $\omega_\mathrm{b}$ and $n_\mathrm{s}$ due to the weak constraining power. This leaves the Hubble constant in eq.~(\ref{eq:method-lcdmparameters}) as final parameter to characterize a \LCDM{} cosmology. Its measurement is informed by (a) the growth rate and (b) a coordinate and amplitude rescaling if distance units are $h/\mpc$. There is no information from the clustering amplitude due to the perfect degeneracy between $D(a)$ and $\clusteringamplitude$. The coordinate rescaling (b) is not a physical effect  \cite{Sanchez:2020vvb}, and the main source of -- geometric -- information on the Hubble rate is absent. In our \LCDM{} analyses, we therefore keep $h$ fixed. The inclusion of Alcock-Paczynski distortions in the forward model will allow to explore the full constraining power of field-level analysis on the lightcone. We leave this for future work.

In addition to \LCDM{}, we consider two cosmologies with a more flexible dependency between the shape of the matter power spectrum and the growth of structure. Such cosmologies arise if gravity differs from General Relativity on large scales. Also modifications to the expansion history affect structure growth, but at a smaller level (c.f. eqs.~(17) and (18) in \cite{Linder:2005in}). In the \gLCDM{} scenario \cite{Linder:2005in}, we express 
\begin{equation}
f(a) = \Omega_\matter(a)^\gamma \quad \mathrm{and} \quad D(a) = \exp\left[\int_{0}^{a} \frac{da'}{a'}\,\Omega_\matter^\gamma(a')\right] \,,
\label{eq:methods-gammamodel}
\end{equation}
and we treat the growth index $\gamma$ as free parameter. \LCDM{} is contained in this cosmology for $\gamma=0.55$. Finally, in the \fLCDM{} scenario, we treat the growth factor $f$ itself as a free parameter. Since $f$ is measured on a single time-slice, we cannot uniquely predict $D(a)$ in this parametrization. Instead we reinterpret  $\clusteringamplitude$ as $\clusteringamplitude D_{f\Lambda\mathrm{CDM}}^2(a) / D^2_{\Lambda\mathrm{CDM}}(a)$. This is possible because $D^2(a)$ and $\clusteringamplitude$ are degenerate to all perturbative orders. This scenario separates information from positions (in $\clusteringamplitude$ and $\omega_\matter$) from peculiar velocities that constrain $f$.

In summary the free parameters of our three cosmologies are
\begin{align}
	\Lambda\mathrm{CDM}:& \quad \alpha\,,~ \omega_\matter\,,\\
	\gamma\Lambda\mathrm{CDM}: & \quad \alpha\,,~ \omega_\matter\,,~ \gamma\,, \\
	f\Lambda\mathrm{CDM}:& \quad \alpha\,,~ \omega_\matter\,,~ f\,. 
\end{align}
where $\alpha \equiv \sqrt{\clusteringamplitude/\clusteringamplitude^\fiducial}$ and $\clusteringamplitude^\fiducial$ is some fiducial value. The ground-truth values and priors of all cosmological parameters are summarized in table~\ref{tab:parameters}.

\begin{table}
	\centering
\begin{tabular}{c|c|c|c}
\textbf{parameter} &  \textbf{definition} & \textbf{fiducial value} & \textbf{prior} \\
\hline \hline
 \multicolumn{4}{l}{{cosmology}} \\
 \hline
$\alpha$ & $\sqrt{\clusteringamplitude/\clusteringamplitude^\fiducial}$  & 0.995814 & $\mathcal{U}\left(0.5, 2.0\right)$ \\ 
$\omega_\matter$ & matter density & 0.1430142 & $\mathcal{U}\left(0.05, 0.40\right)$ \\
$\gamma$ & eq.~(\ref{eq:methods-gammamodel})& 0.55 & $\mathcal{U}\left( -1.45, 2.55 \right)$ \\
$f$ & eq.~(\ref{eq:mathods-growthrate}) & 0.76115 & $\mathcal{U}\left(0.3, 2.0 \right)$ \\
\hline \hline
\multicolumn{4}{l}{{stochasticity}} \\
\hline
$\noiseamplitude$ & eq.~(\ref{eq:methods-noisepowerspectrum}) & 0.41401 & $\mathcal{U}\left(0.32, 10 \right)$ \\
\hline \hline
\multicolumn{4}{l}{{second order bias}} \\
\hline
$b_\sigma$ & \multirow{4}{*}{\makecell{appendix A\\ of \cite{Stadler:2024fui}}} & -0.83645 & $\mathcal{N}\left(0, 10\right) \times \mathcal{U}\left(-100, 100\right)$ \\
$b_{\sigma^2}$ & & -0.18876 & $\mathcal{N}\left(0, 10\right) \times \mathcal{U}\left(-100, 100\right)$ \\
$b_{\mathrm{tr}\left[\left(M^{(1)}\right)^2\right]}$ & & 0.051907 & $\mathcal{N}\left(0, 10\right) \times \mathcal{U}\left(-100, 100\right)$ \\
$b_{\nabla^2\sigma}$ & & 0.0 & $\mathcal{N}\left(0, 10\right) \times \mathcal{U}\left(-100, 100\right)$ \\
$\beta_{\partial_{||}\sigma}$ & eq.~(C.1) of \cite{Stadler:2024aff} & 29.535 & $\mathcal{N}\left(0, 10\right) \times \mathcal{U}\left(-100, 100\right)$\\
\hline \hline
\multicolumn{4}{l}{{third order bias}} \\
\hline
$b_{\sigma^3}$ & \multirow{4}{*}{\makecell{appendix A\\ of \cite{Stadler:2024fui}}} & 0.0 & $\mathcal{N}\left(0, 10\right) \times \mathcal{U}\left(-100, 100\right)$ \\
$b_{\sigma\, \mathrm{tr}\left[\left(M^{(1)}\right)^2\right]}$ & & 0.0 & $\mathcal{N}\left(0, 10\right) \times \mathcal{U}\left(-100, 100\right)$ \\
$b_{\mathrm{tr}\left[\left(M^{(1)}\right)^3\right]}$ & & 0.0 & $\mathcal{N}\left(0, 10\right) \times \mathcal{U}\left(-100, 100\right)$ \\
$b_{\mathrm{tr}\left[M^{(1)} M^{(2)} \right]}$ & & 0.0 & $\mathcal{N}\left(0, 10\right) \times \mathcal{U}\left(-100, 100\right)$ \\
\hline
\end{tabular}
\caption{Free parameters of the inference with fiducial values and priors.}
\label{tab:parameters}
\end{table}

\subsection{Field-level forward model and analysis}
\label{sec:methods-fieldlevel}

Our forward model of the redshift-space galaxy density $\tracerdensityRSD$ is implemented in \LEFTfield{} \cite{Stadler:2023hea, Stadler:2024aff, Schmidt:2020ovm, Stadler:2024fui}. The computation starts from a white-noise field $\hat{s}$, which is resolved up to some initial cut-off $\Lambda$ (and represented on a cubic grid with resolution $N_\mathrm{G,ini}$). The linear density field follows by rescaling with the linear power spectrum,
\begin{equation}
\delta_\linear\left(\vec{k}, z\right) = \sqrt{ \left(\frac{N_\mathrm{G,ini}}{L_\mathrm{box}}\right)^3  P_\linear\left(k, z\right)} \quad \hat{s}\left(\vec{k}\right) \,,
\label{eq:methods-deltaini}
\end{equation}
where $L_\mathrm{box}$ refers to the size of the simulation box. In previous analyses with \LEFTfield{}, the free parameters considered were $\sigma_8$ \cite{Schmidt:2020viy, Schmidt:2020tao, Nguyen:2024yth, Beyond-2pt:2024mqz, Stadler:2024fui}, the BAO scale \cite{Babic:2022dws, Babic:2024wph, Babic:2025fgv} and the growth rate $f$ \cite{Stadler:2023hea, Stadler:2024aff}. The latter does not enter the linear matter power spectrum, and the former two can be implemented via a template rescaling. For this work, \LEFTfield{} is interfaced with the Boltzmann code \CLASS{} \cite{Blas:2011rf}, and we recompute the linear power spectrum on the fly for the current values of the cosmological parameters. 

The higher-order gravitational dynamics are based on third-order Lagrangian Perturbation Theory (LPT) \cite{Schmidt:2020ovm, Rampf:2012up, Zheligovsky:2013eca, Matsubara:2015ipa}. For every location on the initial grid $\vec{q}$, we compute the displacement that combines the gravitational evolution $\vec{s}$ and the peculiar velocity $\vec{v}_\tracerdet$,
\begin{equation}
\vec{\tilde{x}}\left(\vec{q}, \tau\right) = \vec{q} + \vec{s}\left(\vec{q}, \tau\right) + \mathcal{H}^{-1} \left(\vec{v}_\tracerdet \cdot \vec{\hat{n}} \right) \vec{\hat{n}}\,,
\label{eq:methods-displacement}
\end{equation}
to go to redshift space positions $\vec{\tilde{x}}$ in a single step. Here, $\mathcal{H} = aH$ is the reduced Hubble rate and we assume a constant line-of-sight direction $\vec{\hat{n}}$ that is aligned with one of the grid axes. The displacement depends on cosmology through the initial conditions, the growth rate, and the growth factor; we compute the latter two on the fly from eq.~\eqref{eq:methods-grothfactor} or eq.~\eqref{eq:methods-gammamodel}. The redshift-space matter density $\tilde{\delta}_\matter$ follows from displacing an ensemble of particles and assigning their distribution to a grid. For details on the implementation see \cite{Stadler:2024fui, Stadler:2024fui} whose recommendation on the numerical setup we follow.

Applying the displacement from gravitational evolution and redshift-space distortions in a single step implies a Lagrangian basis for the bias expansion. The bias operators are implemented by a weighted displacement, where the weight corresponds to the operator value in the Lagrangian frame,
\begin{equation}
\tilde{\op}\left(\vec{\tilde{x}}\right) = \frac{\left[1 + \delta\left(\vec{\tilde{x}}\right)\right]\, \op\left(\vec{\tilde{x}}\left[\vec{q}\right]\right)}{1 + \partial_{||} \vec{v}_\tracerdet \left(\vec{\tilde{x}}\right)/\mathcal{H}} \,.
\end{equation}
We also allow for a bias between the matter velocity $\vec{v}$ and the tracer velocity $\vec{v}_\tracerdet$, that is already implicit in the displacement in eq.~\eqref{eq:methods-displacement}. The equivalence principle ensures that bias terms in the velocity are derivative suppressed, and the tracer velocity is
\begin{equation}
\vec{v}_\tracerdet = \vec{v} + \beta_{\partial_{||}\sigma}\, \partial_{||} \sigma + \ldots \,,
\end{equation}
The galaxy density in redshift space is finally constructed from the displaced operators as \cite{Stadler:2024aff}
\begin{equation}
\tracerdensityRSD\left(\vec{\tilde{x}}\right) = \tilde{\delta}_\matter + \sum_\op b_\op \,\op\left(\vec{\tilde{x}}\right) \,.
\label{eq:methods-vtracerdet}
\end{equation}
For our baseline analysis we perform the bias expansion up to second order, and we consider an extended scenario with third order bias. In both cases, we include the leading-order higher-derivative contribution $\nabla^2\sigma$ and the leading velocity operator as given by eq.~\eqref{eq:methods-vtracerdet}. In total, there are five bias coefficients in the baseline second order bias expansion and nine coefficients at third order. They are summarized with their priors in table~\ref{tab:parameters}.

For the galaxy stochasticity, we consider the leading-order white-noise contribution. In principle, the transformation to redshift-space leads to anisotropic noise \cite{Cabass:2020jqo}, which we neglect. As explained in section~\ref{sec:data}, our data is consistent with this assumption, such that we do not expect biases from it. The likelihood then is given by eq.~\eqref{eq:intro-likelihood}, with the replacements $\delta_\tracer \rightarrow \tilde{\delta}_\tracer$, and $\delta_\tracerdet \rightarrow \tilde{\delta}_\tracerdet$. Bias coefficients enter the likelihood quadratically, and can be marginalized over analytically \cite{Elsner:2019rql}. Analytic marginalization does not work for the velocity-dependent bias, whose coefficient is always explored numerically. We draw samples from the marginalized posterior for numerical efficiency \cite{Kostic:2023arx}. Where constraints on the bias coefficients are of interest, they are obtained in retrospect from the conditional posterior that we derive in appendix~\ref{appx:parsampling}. The variance of the likelihood, $\noiseamplitude^2$, is unknown a priori and also inferred. It relates to the noise power spectrum as
\begin{equation}
P_\epsilon = \noiseamplitude^2 \left(L_\mathrm{box}/N_\mathrm{G}\right)^3 \,,
\label{eq:methods-noisepowerspectrum}
\end{equation} 
where $N_\mathrm{G}$ is the minimum grid size to represent all modes up to the likelihood cutoff $k_\mathrm{max}$. Throughout this paper, we choose the likelihood cutoff as $k_\mathrm{max} = \Lambda = 0.14\,h/\mpc$ and set the box size to $L_\mathrm{box} = 2\,\gpc/h$ yielding $N_\mathrm{G}= 90^3$. The most important parameters of our configuration are summarized in table~\ref{tab:metaparameters}.

Summing up, there are $729.004$ free parameters in the \LCDM{} analysis, and one more for the extended cosmologies. We sample from the high-dimensional space of initial conditions with Hamiltonian Monte Carlo (HMC, \cite{Neal:2011mrf}). In a block-sampling approach, the HMC updates are interlaced with slice-sampling  \cite{2000physics...9028N} of the cosmological parameters, noise amplitude, and the velocity bias coefficient. The sampling procedure is implemented in \LEFTfield{}, and details of the initialization, stopping criteria and convergence are presented in appendix~\ref{appx:hmc}. We collect at least 100 effective samples and aim for a Gelman-Rubin criterion $R-1 \leq 0.05$ for all parameters of interest\footnote{The only exception is $\noiseamplitude$ in the \LCDM{} analysis with third order bias, where a very long correlation length leads to a slightly reduced effective sample size and increased $R-1$, see appendix \ref{appx:hmc} for details.}.

\subsection{Power- and Bispectrum forecast}
\label{sec:methods-fisher}

We perform a Fisher forecast \cite{Fisher:1935,Tegmark:1997rp} to predict parameter constraints from the power- and bispectrum. Our goal is to compare field-level analysis to power- and bispectrum on the exactly same mock data set and with identical modeling choices for both analyses. The Fisher matrix is given by
\begin{equation}
F_{ij} = \frac{\partial \left\langle\vec{d} \right\rangle}{\partial \vartheta_i} \Sigma^{-1} \frac{\partial \left\langle\vec{d} \right\rangle}{\partial \vartheta_j} + \left(\Sigma_\mathrm{prior}^{-1} \right)_{ij} \,,
\end{equation}
where $\vec{d}$ is the data vector, $\left\langle\vec{d} \right\rangle$ its expectation over the likelihood, and $\Sigma$ its covariance. Bias coefficients, noise amplitude and cosmology parameters (i.e. the parameters given in table~\ref{tab:parameters}) are summarized in $\vartheta$, and $\Sigma_\mathrm{prior}$ is the prior covariance which is diagonal in our case.

Our data vector comprises the power spectrum multipoles $P^{(l)}(k)$, $l=0,2,4$ and the bispectrum monopole $B^{(0)}(k_1, k_2, k_3)$ with linear binning in $k$ of width $\Delta k = 4\pi/L_\mathrm{box}$ (twice the fundamental mode of the box). This corresponds to $22$ bins in the power spectrum, $1199$ triangles in the bispectrum and a total data vector of size $1265$. We compute possible data realizations by adding random noise draws to the  \LEFTfield{} deterministic prediction, $\tilde{\delta}_\tracer = \tilde{\delta}_\tracerdet + \epsilon_\tracer$, and we estimate the data vector as \cite{Sefusatti:2015aex, Scoccimarro:1997st, Scoccimarro:2015bla}
\begin{align}
\hat{P}^{(l)}(k) &= \frac{2l+1}{N_k \, L_\mathrm{box}^3} \sum_{\vec{q}\in k} \mathcal{L}_l\left(\hat{\vec{q}}\cdot \hat{\vec{n}}\right) \left|\tilde{\delta}_\tracer\left(\vec{q}\right)\right|^2 \,, \\
\hat{B}^{(0)}\left(k_1, k_2, k_3\right) &= \frac{1}{N_t \ L^3_\mathrm{box}} \,  \sum_{\vec{q}_1, \vec{q}_2, \vec{q}_3 \in \mathrm{Tri}_{123}} \tilde{\delta}_\tracer\left(\vec{q}_1\right) \, \tilde{\delta}_\tracer\left(\vec{q}_2\right) \, \tilde{\delta}_\tracer\left(\vec{q}_3\right) \,,
\end{align}
where $\mathcal{L}_l$ refers to the Legendre polynomial of $l$-th order, $k_1 \leq k_2 \leq k_3$, and $\mathrm{Tri}_{123}$ is the set of triangles that can be formed such that $|\vec{k}_i - \vec{q}_i| < \Delta k /2$ and $\vec{q}_1 + \vec{q}_2 + \vec{q}_3 = 0$. This also includes ``open triangles'' \cite{Ivanov:2021kcd} for which the triangle formed by the bin centers does not close. The estimators are implemented via the integral representation of the Dirac delta, and we super-sample all grids by a factor of two to prevent aliasing below $k_\mathrm{max}$.

We estimate derivatives of the data vector from finite differences, i.e. two \LEFTfield{} simulations with identical initial conditions and small variations in one parameter at a time. To obtain the expected derivatives that enter the Fisher matrix, we average these estimates over an ensemble of initial conditions. The covariance is estimated from an ensemble of simulations with different initial conditions and noise realizations at the true cosmology, and we also account for off-diagonal elements. Details on the number of simulations and the step size for the numerical difference are provided in appendix~\ref{appx:fisher}. We estimate that the limits on cosmological parameters are accurate to at least $2-3\,\%$.

\section{Data}
\label{sec:data}

\begin{table}
	\centering
\begin{tabular}{c|c|c}
\textbf{parameter} & \textbf{meaning}& \textbf{value} \\
\hline
\hline
\multicolumn{3}{l}{fiducial cosmology (the \texttt{AbacusSummit} \cite{Maksimova:2021ynf} \texttt{c000} cosmology)}\\
\hline
$\omega_\matter$ & matter density &  0.1200\\
$\omega_\mathrm{b}$ & baryon density & 0.02237\\
$\clusteringamplitude$ & primordial amplitude &  $2.0830 \times 10^{-9}$\\
$n_\mathrm{s}$ & primordial tilt & 0.9649 \\
$h$ & Hubble constant in units of $100\,\mathrm{km}/\left(\mathrm{s}\,\mpc\right)$ &  0.6736\\
\hline
\hline
\multicolumn{3}{l}{fiducial mock properties} \\
\hline
$z$ & redshift & 0.5 \\
$P_\epsilon$ & noise power spectrum & $1.9 \times 10^3 \, \left(\mpc/h\right)^3$\\
$b_1$ & $\simeq 1 - b_\sigma$, linear bias & $\simeq 1.84$\\
$V_\mathrm{eff}$ & effective volume at $k=0.14\,h/\mpc$ & $5.6\,\left(\gpc/h\right)^3$\\
\hline
\hline
\multicolumn{3}{l}{noisy mock properties (parameters not provided are identical to the fiducial mock)} \\
\hline
$P'_\epsilon$ & noise power spectrum & $3.8 \times 10^3 \, \left(\mpc/h\right)^3$\\
$V'_\mathrm{eff}$ & effective volume at $k=0.14\,h/\mpc$ & $4.1\,\left(\gpc/h\right)^3$ \\
\hline
\hline
\multicolumn{3}{l}{scales used in the analysis} \\
\hline
$L_\mathrm{box}$ & size of the cubic simulation box & $2\,\gpc/h$ \\
$\Lambda$ & Initial conditions cut-off & $0.14\,h/\mpc$ \\ 
$k_\mathrm{max}$ & likelihood cut-off & $0.14\,h/\mpc$ \\
$N_\mathrm{G}$ & grid size per dimension ($\delta_\linear$ and $\delta_\tracerdet$) & 90 \\
$\Delta k$ & bin width in power- and bispectrum & $4\pi/L_\mathrm{box}$ \\
\hline 
\end{tabular}
\caption{Summary of the key parameters characterizing the mock data set and analysis.}
\label{tab:metaparameters}
\end{table}

\begin{figure}
\centering
\includegraphics[]{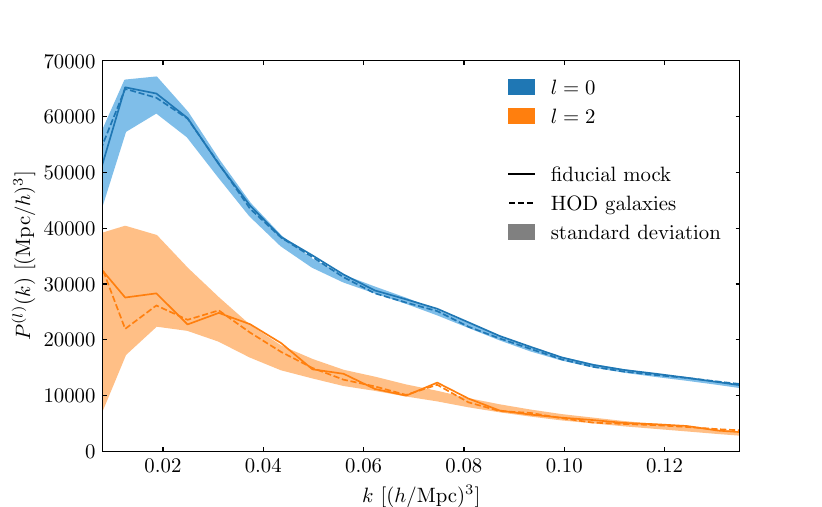}
\caption{Power spectrum of the fiducial mock (solid), which was generated from the perturbative forward model. Bias coefficients and noise amplitude for the mock have been obtained from a fit to HOD galaxies which closely resemble DESI LRGs; their power spectrum is shown by the dashed lines. Shaded regions indicate the mean-centered standard deviation of the power spectrum from $4\times 10^{5}$ \LEFTfield{} simulations with independent initial conditions.}
\label{fig:powerspectra}
\end{figure}

We use the \LEFTfield{} forward model to generate mock data in which we can control noise properties and higher-order bias. In particular non-Gaussian noise can lead to parameter miss-estimation, so-called ``sigma collapse'' \cite{Akitsu:2025boy,Rubira:2025rqo}. While it is possible to include higher-order noise in the field-level forward model, this also significantly increases the computational demands of an inference \cite{Rubira:2025rqo}. For the purpose of forecasting the constraining power of field-level analysis, focusing on the leading order noise contribution allows us to analyze several cosmological scenarios and to demonstrate their convergence.

To generate mock data that closely resemble observed galaxy samples, we determine bias coefficients and the stochastic amplitude from HOD-based mocks that have been developed for DESI \cite{Yuan:2021izi, Yuan:2023ezi}. In particular, we use their baseline HOD model for Luminous Red Galaxies with parameters measured from the one-percent survey\footnote{The HOD model is based on \cite{Zheng:2007zg} and extended by one incompleteness and two velocity bias parameters. For all parameters, we use the central values from Table 3 in \cite{Yuan:2023ezi}.} in the redshift range $0.4 < z < 0.5$ to populate a base simulation of the \texttt{AbacusSummit} suite \cite{Garrison:2021lfa, Maksimova:2021ynf, Hadzhiyska:2021zbd}\footnote{We pick the cleaned halo catalogs from the \texttt{c000\_ph000} simulation.}. We assign these galaxies to a grid and measure the bias coefficients with \LEFTfield{} at fixed cosmology and initial conditions, keeping otherwise the same analysis configuration as described in section~\ref{sec:methods-fieldlevel}. Finally, we use the measured coefficients (see table~\ref{tab:parameters}) to create the ``fiducial mock'' from the same initial conditions as the original simulations with additive Gaussian noise. We create a second, ``noisy mock'' in which we double the noise power spectrum, keeping all other parameters identical. Key properties of the mocks are summarized in table~\ref{tab:metaparameters}. In figure~\ref{fig:powerspectra}, we compare the power spectrum monopole and quadrupole of the fiducial mock to the original HOD galaxies, finding very good agreement up to the cut-off $k_\mathrm{max} = 0.14\,h/\mpc$.

As argued in the introduction, the performance of field-level analysis depends on the data. We therefore put our mocks in context with previous studies that compare field-level analysis to the power- and bispectrum, namely the \pnameBeyond{} \cite{Beyond-2pt:2024mqz}, \pnameNguyen{} \cite{Nguyen:2024yth}, and \pnameAkitsu{} \cite{Akitsu:2025boy}. Our box size, $L_\mathrm{box}=2\,\gpc$, is identical to these works. The tracers of \pnameNguyen{} are halos in the restframe of N-body simulations; they use one population at redshift $z=0.5$ with number density $\bar{n}=1.3\times 10^{-3}\,\left(h/\mpc\right)^3$ and another at $z=1.03$ with $\bar{n}=3.6\times 10^{-3}\,\left(h/\mpc\right)^3$. The tracers for field-level analysis in the \pnameBeyond{} are HOD-based galaxies, again in the restframe, at $z=1.0$ with $\bar{n}=4.5\times 10^{-4}$. Closest in spirit to this work are the redshift-space snapshots of \pnameAkitsu{}, which are generated from their perturbative forward model at $z=0.5$ and have additive Gaussian noise of $P_\epsilon=3.0\times 10^3 \left(\mpc/h\right)^3$. Our fiducial mock is somewhat intermediate to previous studies, it corresponds to a lower number density than in \pnameNguyen{} and slightly higher one than in the \pnameBeyond{}. We reiterate that the fiducial mock parameters closely resemble DESI LRGs to give a realistic comparison between field-level and power- and bispectrum analyses. Our noisy mock, on the other hand, exhibits a higher noise power spectrum or a lower number density (assuming Poisson counts) than all previous works.

\section{Results and discussion}
\label{sec:results}

By performing field-level analysis of the mock data sets and comparing it to a Fisher forecast of the power- and bispectrum in an identical configuration, we estimate how much more information field-level analysis can extract on cosmological parameters. The additional information can originate in principle from higher-order correlations and from higher-order bispectrum multipoles. Nevertheless, bispectrum multipoles with $l>0$ have previously been found to impact  cosmological parameters only marginally \cite{Ivanov:2023qzb}. As argued in section~\ref{sec:methods-fisher}, we expect the Fisher forecast to accurately reflect the constraining power from power- and bispectrum, and at worst give it too much credit. In this sense our estimates are conservative. We first discuss results for a \LCDM{} cosmological model in subsection~\ref{sec:results-LCDM}, and we move to extended cosmologies in subsection~\ref{sec:results-extensions}. All analyses and their results are summarized in table~\ref{tab:results}.

\begin{table}
	\centering
	\begin{tabular}{c c|c|c|c}
		& parameter & \makecell{field-level\\ half 68\,\% interval} & \makecell{power- \& bispectrum\\ standard deviation} & \makecell{\bf field-level\\ \bf improvement}	\\
		\hline
		\hline
		\multicolumn{5}{l}{\textbf{\LCDM{} analyses} (section~\ref{sec:results-LCDM})}  \\
		\hline
		\multirow{3}{*}{\makecell{fiducial mock,\\ 2nd order bias\\ \footnotesize{figure~\ref{fig:results-baseline} and \ref{fig:results-violin}}}\hspace{.1cm}} 
		& $\alpha$ & $1.6\times 10^{-2}$ & $3.3 \times 10^{-2}$ & 2.0 \\
		& $\omega_\matter$ & $1.9 \times 10^{-3}$ & $2.0 \times 10^{-3}$ & $1.0$\\
		& $\noiseamplitude$ & $8.0 \times 10^{-3}$ & $2.2 \times 10^{-2}$ & $2.7$\\
		\hline
		\multirow{3}{*}{\makecell{fiducial mock,\\ 3rd order bias\\ \footnotesize{figure~\ref{fig:results-lcdm-extensions}} }\hspace{.1cm}} 
		& $\alpha$ & $1.7 \times 10^{-2}$ & $3.6 \times 10^{-2}$ & $2.1$\\
		& $\omega_\matter$ & $2.1 \times 10^{-3}$ & $2.5\times 10^{-3}$ & $1.2$ \\
		& $\noiseamplitude$ & $1.2\times 10^{-2}$ & $3.1 \times 10^{-2}$ & $2.6$ \\
		\hline
		\multirow{3}{*}{\makecell{noisy mock,\\ 2nd order bias\\ \footnotesize{figure~\ref{fig:results-lcdm-extensions}}}\hspace{.1cm}} 
		& $\alpha$ & $2.1\times 10^{-2}$ & $3.6\times 10 ^{-2}$ & $1.7$ \\
		& $\omega_\matter$ & $2.1\times10^{-3}$ & $2.2\times10^{-3}$ & $1.0$ \\
		& $\noiseamplitude$ & $6.9\times10 ^{-3}$ & $1.2\times10^{-2}$ & $1.7$\\
		\hline
		\hline
		\multicolumn{5}{l}{\textbf{modified growth histories}, fiducial mock, 2nd order bias (section~\ref{sec:results-extensions})}  \\
		\hline
		\multirow{4}{*}{\makecell{\gLCDM{} \\ \footnotesize{figure~\ref{fig:results-beyond-lcdm} }}\hspace{.1cm}} 
		& $\alpha$ & $1.7\times 10^{-2}$ & $4.4 \times 10^{-2}$ & 2.6 \\
		& $\omega_\matter$ & $1.9 \times 10^{-3}$ & $2.1 \times 10^{-3}$ & $1.1$\\
		& $\gamma$ & $7.4\times10^{-2}$ & $1.4\times10^{-1}$ & $1.9$\\
		& $\noiseamplitude$ & $7.0 \times 10^{-3}$ & $2.2 \times 10^{-2}$ & $3.1$\\
		\hline
		\multirow{4}{*}{\makecell{\fLCDM{} \\ \footnotesize{figure~\ref{fig:results-beyond-lcdm} and \ref{fig:results-violin}}} \hspace{.1cm}} 
		& $\alpha$ & $2.1 \times 10^{-2}$ & $6.5 \times 10^{-2}$ & $3.1$ \\
		& $\omega_\matter$ & $2.0\times 10^{-3}$ & $2.1\times10^{-3}$ & 1.1 \\
		& $f$ & $2.6\times10^{-2}$ & $5.4\times10^{-2}$ & $2.0$ \\
		& $\noiseamplitude$ & $7.1\times10^{-3}$ & $2.2\times10^{-2}$ & 3.1\\
	\end{tabular}
	\caption{Measurement precision for cosmological parameters in field-level analysis and for the power- and bispectrum, where $\alpha = \sqrt{\clusteringamplitude / \clusteringamplitude^\mathrm{fid.}}$. The table summarizes all analysis configurations explored in this paper. Power- and bispectrum limits are based on a Fisher forecast and we report the $1\sigma$ width.  For field-level results, we report half the size of the of the 68\,\% credible interval, which is equivalent to the $1\sigma$ width for Gaussian posteriors. The field-level improvement in the final column is calculated as the ratio of the power- and bispectrum standard deviation over the field-level interval.}
	\label{tab:results}
\end{table}

\subsection{\LCDM{}}
\label{sec:results-LCDM}

\begin{figure}
	\centering
	\includegraphics[]{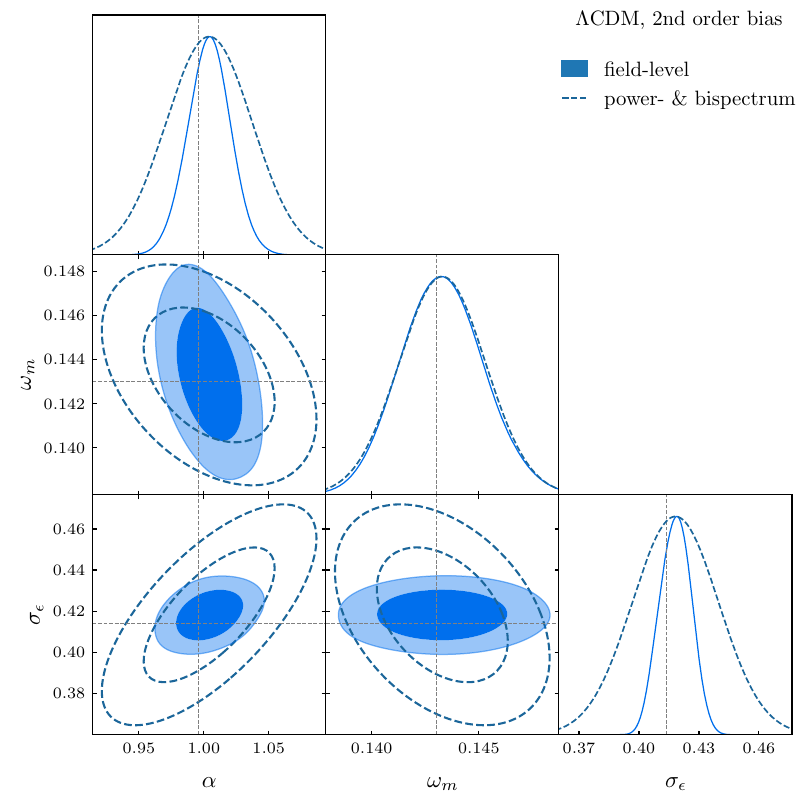}
	\caption{Comparison between constraints on cosmological parameters from field-level analysis (filled contours) and from the power- and bispectrum (dashed lines) on the fiducial mock data set under a \LCDM{} cosmology with second order bias. The power- and bispectrum constraints are estimated from a Fisher forecasts and contours are centered on the mean values found in field-level analysis. Field-level analysis considerably improves the measurement of the clustering- and noise amplitude, as can be also seen from the one-dimensional intervals quoted in table~\ref{tab:results}.}
	\label{fig:results-baseline}
\end{figure}

For our baseline scenario, we consider the fiducial mock, a \LCDM{} cosmology and the second order bias expansion. The cosmological constraints are summarized in figure~\ref{fig:results-baseline} and table~\ref{tab:results}. The true values of all cosmological parameters lie inside the 68\,\% contour of the field-level analysis. The mock data is generated from the same forward model as used in the analysis, such that no biases are expected. Still, the result points to the absence of prior volume effects which can move the inferred parameter intervals away from the ground truth even in the absence of model misspecification. The contours for the power- and bispectrum analysis are based on a Fisher forecast and have been centered on the field-level mean.

\begin{figure}
	\centering
	\includegraphics[]{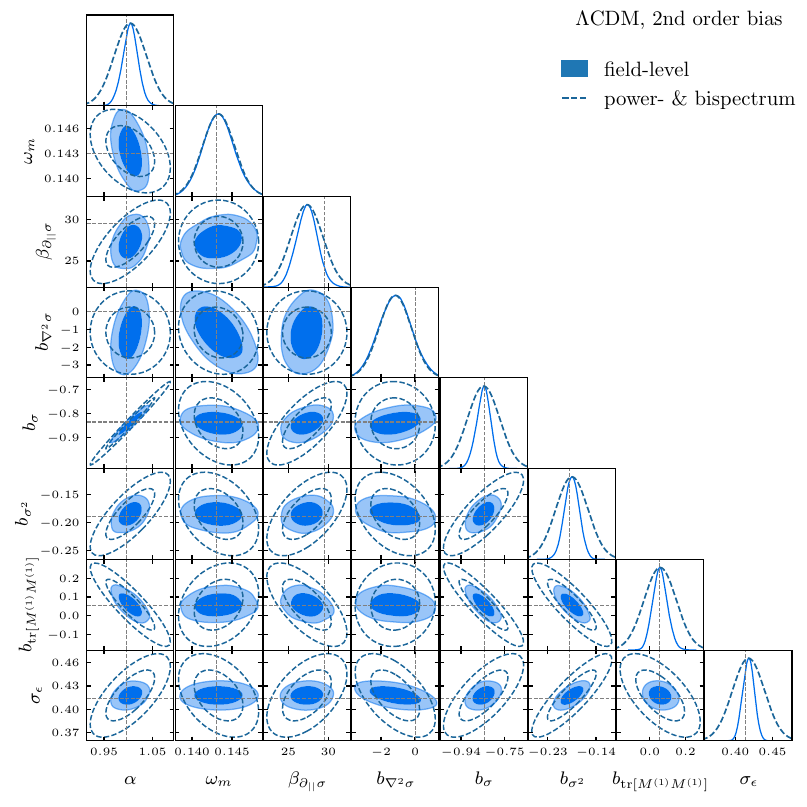}
	\caption{Constraints in the combined space of cosmological parameters and bias coefficients for the baseline \LCDM{} analysis on the fiducial mock (see also figure~\ref{fig:results-baseline}). Filled contours represent the field-level results and dashed lines the power- and bispectrum. }
	\label{fig:appx-parsampling-lcdm}
\end{figure}

Comparing the constraining power of the two analyses, we first note their agreement on the matter density $\omega_\matter$. This parameter is determined precisely from the shape of the linear power spectrum and higher-order information has little impact on its precision. The situation is different for the clustering amplitude $\alpha \propto\sqrt{\clusteringamplitude}$, which is determined twice as precisely at the field level. Figure~\ref{fig:results-baseline} reveals a mild degeneracy with the noise amplitude $\sigma$ that is broken effectively in the field-level analysis. That is, field-level information helps to better disentangle signal and noise, and correspondingly also the precision at which the noise amplitude can be determined improves significantly.

Nevertheless, degeneracy with the noise is not the only reason why field-level analysis improves the precision of the primordial amplitude. To get a more complete picture, figure~\ref{fig:appx-parsampling-lcdm} shows parameter constraints in the extended space of cosmological parameters and bias coefficients. Several coefficients exhibit a degeneracy with $\alpha$, most notably the linear bias, $b_\sigma$. At first glance, this seems to contradict the arguments made in subsection~\ref{sec:methods-cosmology}, i.e. that primordial amplitude and bias can be disentangled because redshift-space distortions allow to measure the parameter combinations $b_1 \sigma_8$ and $f \sigma_8$ from the power spectrum monopole and quadrupole at linear order and $f$ is determined by $\omega_\matter$ (we remind the reader that $h$ is fixed). An intuitive reason for the remaining bias-amplitude degeneracy might be the different measurement precisions for the power spectrum monopole and quadrupole (see figure~\ref{fig:powerspectra}) and their non-trivial cross-correlation. In fact, many earlier works on full-shape analyses of power spectrum multipoles observe a similar degeneracy, see e.g. \cite{Nishimichi:2020tvu, Donald-McCann:2023kpx}. As a consequence, higher-order information contributes significantly to the constraining power, and field-level analysis improves significantly over power- and bispectrum.

\begin{figure}
\centering
\includegraphics[]{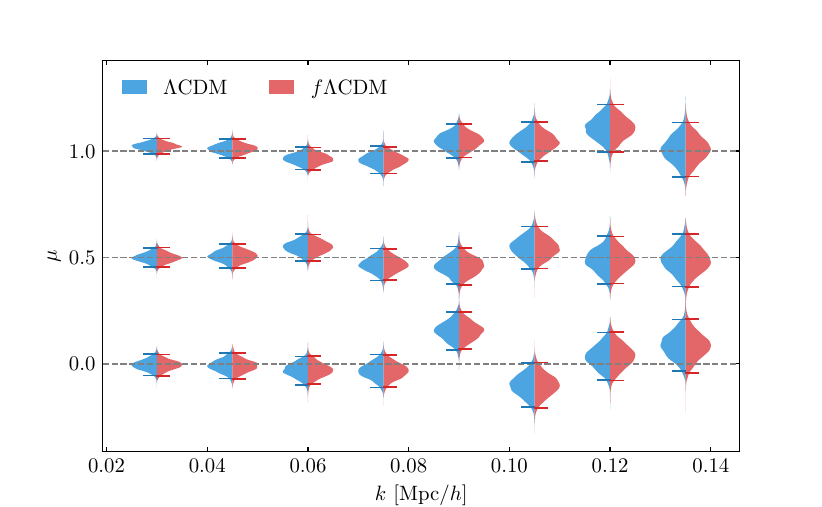}
\caption{Posterior distributions for some individual modes in the initial conditions $\hat{s}$, see eq.~\eqref{eq:methods-deltaini} for definition. We consider two different cosmological models, \LCDM{} and an extension where the growth rate $f$ is treated as free parameter (\fLCDM{}). Modes are selected by their norm $k$ and their angle with the line-of-sight $\mu = \vec{k} \cdot \hat{n} / k$, where the two components of $\vec{k}$ perpendicular to the line of sight have equal magnitude. The posteriors are shown relative to the ground truth (dashed lines), and different values of $\mu$ are offset by four times the prior standard deviation. Vertical bars indicate the percentiles corresponding to the $2\sigma$ range. We represent $\hat{s}$ in the Hartley convention.}
\label{fig:results-violin}
\end{figure}

In addition to the cosmological parameters and bias coefficients, also the initial phases $\hat{s}$ are free parameters of the field-level analysis. Their posterior is shown for a selection of modes in figure~\ref{fig:results-violin}. We find that the initial conditions are overall well constrained. Their precision increases for lower wavenumbers and, to a lesser extent, for directions parallel to the line of sight. Both trends can be understood from the linear power spectrum approximation,
\begin{equation}
	P{\tracer\tracer}\left(k, \mu\right) \simeq \left(b_1 + \mu^2 f\right)^2 P_\linear(k) \,,
\end{equation}
where $\mu = \vec{k}\cdot\hat{n}/k$ denotes the angle with the line of sight. Since the noise power spectrum remains constant, large modes parallel to the line of sight exhibit the larges signal-to-noise ratio and are best constrained. More quantitatively, in the linear approximation and at fixed cosmology, the posterior of the initial conditions is given by the analytic Wiener Filter solution \cite{Kostic:2023arx}. In this approximation, all modes are independent and their standard deviation scales as
\begin{equation}
\sigma_\mathrm{WF}\left(k, \mu\right) \propto \left[1 + \frac{P_{\tracer\tracer}\left(k,\mu\right)}{P_\epsilon}\right]^{-\frac{1}{2}} \,.
\end{equation}
Between the peak of the power spectrum and the smallest scale depicted in figure~\ref{fig:results-violin}, $\sigma_\mathrm{WF}$ evolves by a factor $\sim 2.3$, while from $\mu=0$ to $\mu=1$ the increase is by $\sim 1.3 - 1.4$. 

\begin{figure}
\centering
\includegraphics[]{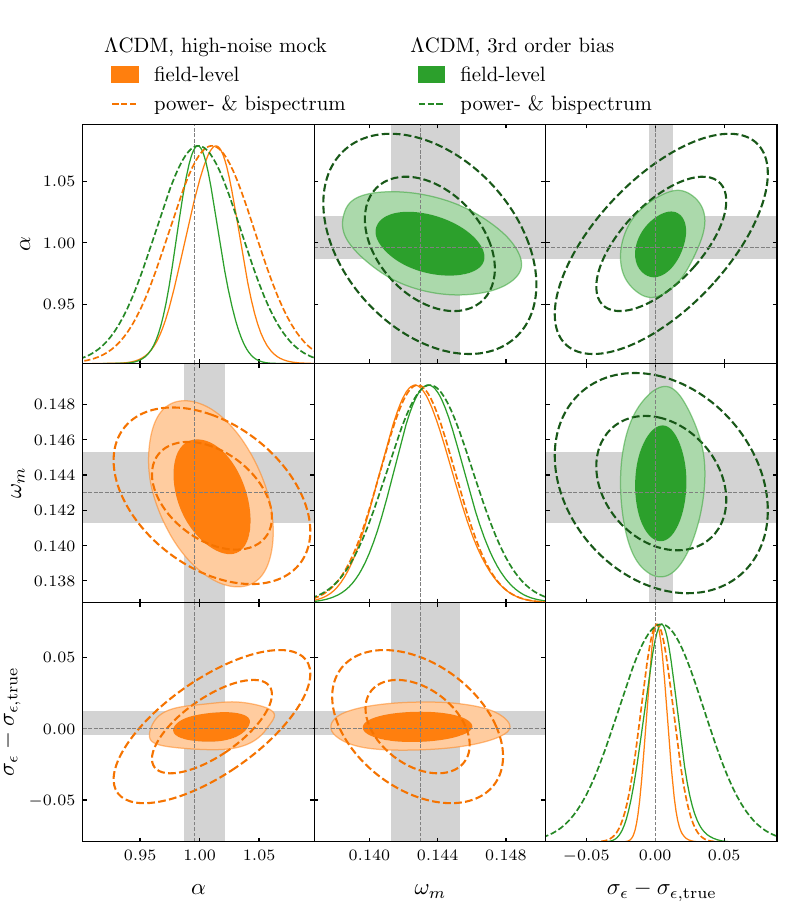}
\caption{Comparison between cosmological constraints from field-level analysis (solid) and power- and bispectrum (dashed) for two extensions of the baseline analysis. The bottom left corner (orange) considers the noisy mock, while the top right corner (green) extends the bias expansion to third order, introducing four additional parameters. Gray bands indicate the one-dimensional 68\,\% interval from the baseline analysis on the fiducial mock (figure~\ref{fig:results-baseline}) for reference. The improvement on clustering- and noise amplitude from field-level analysis remain considerable for both cases, see also table~\ref{tab:results}.}
\label{fig:results-lcdm-extensions}
\end{figure}

All parameter constraints reported for the baseline analysis are more precise than those found in \pnameAkitsu{} \cite{Akitsu:2025boy}. We also find stronger fractional improvements in the field-level precision over the power- and bispectrum. The three main differences with \pnameAkitsu{} are: (a) our choice to fix $h$ (see section~\ref{sec:methods-cosmology}) (b) the higher cut-off $k_\mathrm{max} = 0.14\,h/\mpc$ used in this work versus their $k_\mathrm{max} = 0.10\,h/\mpc$ and (c) the lower noise level of our fiducial mock. While fixing the Hubble constant (a) clearly contributes to tighter absolute constraints, we do not expect it to drive the fractional improvement, rather the opposite. Following the discussion in section~\ref{sec:methods-cosmology} and ignoring unphysical effects from coordinate rescaling, the Hubble constant is mainly constrained from redshift-space distortions. Varying $h$ in the inference leads to more flexibility in the growth rate and partly reinstates the bias-amplitude degeneracy. Field-level analysis breaks the degeneracy more effectively, and we would expect even stronger fractional improvements if $h$ was varied. Indeed, we observe such a trend for extended models with more flexible growth histories in the following subsection. Moving to the cut-off scale (b), a higher value clearly drives both, the absolute and fractional, improvements. The latter because higher-order effects and hence non-Gaussian information are the most pronounced on small scales. To test how strongly the improvement from field-level analysis is driven by the stochasticity of the data (c), we repeat the baseline (\LCDM{}, second order bias) analysis for the noisy mock. Its error power spectrum exceeds the that from previous field-level studies \cite{Akitsu:2025boy, Beyond-2pt:2024mqz, Nguyen:2024yth}. Because the signal power spectrum decreases towards higher wavenumbers, the increased noise first erases information from high-$k$ which benefits field-level analysis particularly. In that sense, results from the noisy mock provide a conservative limit on the precision gains achievable in field-level analysis. 

Results for the noisy mock are summarized in figure~\ref{fig:results-lcdm-extensions} and table~\ref{tab:results}. As it was the case for the fiducial mock, the matter density is equally well constrained by power- and bispectrum and by field-level analysis. The precision of the noise amplitude $\noiseamplitude$ increases in comparison to the fiducial mock for both analyses, but the improvement is stronger for power- and bispectrum. Consequently, the fractional improvement on $\noiseamplitude$ is lessened on the noisy mock. Similarly, the improvement on $\alpha$ is slightly reduced. Nevertheless, field-level analysis increases the precision on the clustering amplitude by a considerable factor of $1.7$ even in the conservative scenario.

Finally, we extend the \LCDM{} analysis of the fiducial mock to third-order bias. Additional degeneracies arise between cosmology and third-order bias coefficients, and we expect the improvement from field-level analysis to be even stronger than in the baseline scenario. Indeed, we find some moderate further improvement on the clustering amplitude, see table~\ref{tab:results} and figure~\ref{fig:results-lcdm-extensions}. For third order bias, also the precision of the matter density is affected by the analysis choice, and field-level analysis leads to a moderate $20\,\%$ improvement. This improvement indicates some degeneracy between the shape of the power spectrum and third-order bias that is broken more effectively at the field level.

\subsection{Modified growth histories}
\label{sec:results-extensions}

In \LCDM{}, the growth rate $f$ is uniquely predicted by $\omega_\matter$ and $h$, with the latter kept fixed in all our analyses. With $\omega_\matter$ determined precisely from the power spectrum shape, redshift-space distortions then break the bias-amplitude degeneracy that persists in the galaxies restframe. At the linear level, this degeneracy breaking can be understood from the simultaneous measurement of $b_1 \sigma_8$ and $f\sigma_8$ (see section~\ref{sec:methods-cosmology}). Nevertheless, the previous section~\ref{sec:results-LCDM} showed how a remaining degeneracy between amplitude and bias is effectively broken by field-level information. Scenarios with modified growth histories, such as the \gLCDM{} and the \fLCDM{} scenario introduced in section~\ref{sec:methods-cosmology}, allow more flexibility between the power spectrum shape and structure growth, so $f$ is no longer uniquely predicted by $\omega_\matter$. These scenarios reinstate, at least in parts, the linear-order degeneracy between clustering amplitude and bias. Therefore, we can expect that field-level analysis provides even stronger improvement. 

\begin{figure}
\centering
\includegraphics[]{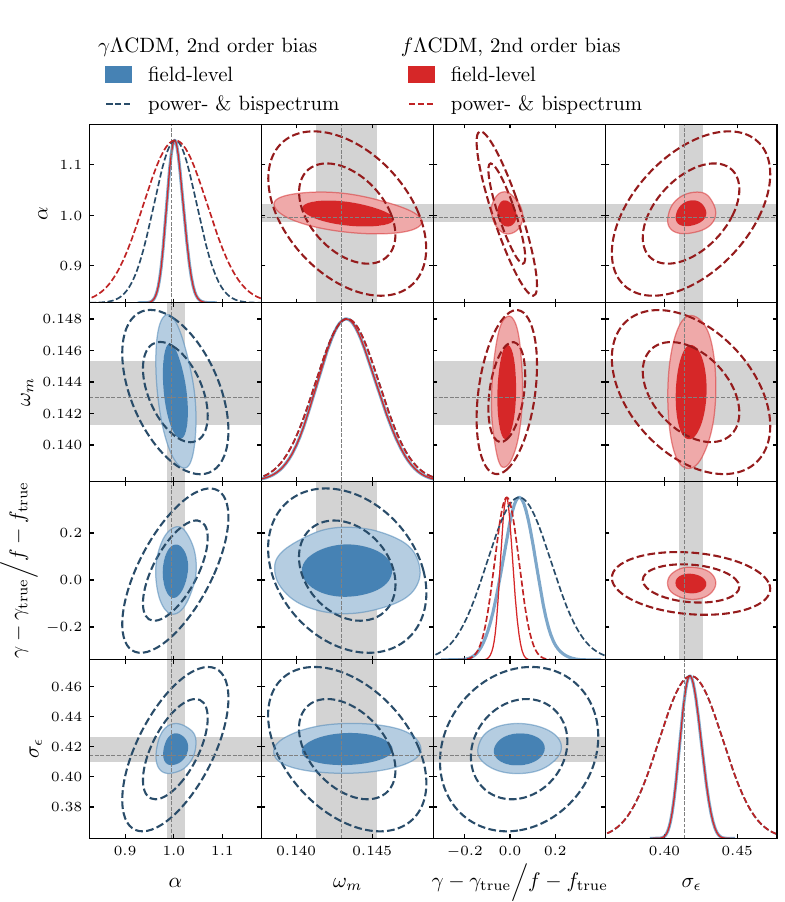}
\caption{Comparison between field-level analysis (solid) and power- and bispectrum (dashed) for two extended cosmologies where the growth history has been modified w.r.t. \LCDM{}. In the bottom left corner (\gLCDM{}, blue), the growth rate is parametrized as $f=\Omega_\matter^\gamma$, while in the top right corner (\fLCDM{}, red), $f$ itself is treated as free parameter, see section~\ref{sec:methods-cosmology} for details. All analyses are for the fiducial mock. Gray bands indicate the one-dimensional standard deviation from the baseline \LCDM{} analysis on the fiducial mock (figure~\ref{fig:results-baseline}) for reference.}
\label{fig:results-beyond-lcdm}
\end{figure}

The results from field-level analysis and from the power- and bispectrum for cosmologies with more flexible growth histories are summarized in figure~\ref{fig:results-beyond-lcdm} and in table~\ref{tab:results}. As for \LCDM{} scenarios with second order bias expansion, the matter density is equally well constrained in both analyses. Constraints on the clustering amplitude are only slightly enlarge compared to \LCDM{} in field-level analysis, but they increase considerably for the power- and bispectrum. Consequently, the improvement from field-level analysis over power- and bispectrum is even more considerable in the extended cosmologies. For the \fLCDM{} analysis, we find gains up to a factor of three. Also the new parameters that describe deviations from \LCDM{} can be measured more precisely at the field level; they improve by roughly a factor of two.

Constraints on some selected modes are shown in figure~\ref{fig:results-violin} for the \LCDM{} and \fLCDM{} cosmologies. There is little difference between \LCDM{} and the extended \fLCDM{} scenario even for modes parallel to the line-of-sight. A similar behavior is observed for the \gLCDM{} cosmology and for \LCDM{} with third order bias. Of all scenarios considered, only \LCDM{} on the noisy mock broadens the phase posterior. We can interpret this in a hierarchical Bayesian inference framework, where cosmological parameters act as hyper-priors on the phases by defining plausible initial power spectra and gravitational dynamics. More freedom in either of these from a more flexible cosmology or bias model apparently does not translate into wider phase posteriors, indicating they are driven by the data through the likelihood.

\section{Conclusions and outlook}
\label{sec:conclusions}

This work quantifies the improvement in the precision of cosmological parameters from field-level analysis over the power- and bispectrum. It extends previous works with \LEFTfield{} \cite{Beyond-2pt:2024mqz, Nguyen:2024yth} to redshift space and a more flexible cosmology model, and it complements related analysis with differing mock and forward model choices \cite{Akitsu:2025boy}. The data sets considered for this work are generated from the forward model and closely resemble the Luminous Red Galaxy sample from DESI (section~\ref{sec:data}). All results are summarized in table~\ref{tab:results}. 

The main finding of this work is a substantial improvement in the precision at which the clustering amplitude $D(a)\,\sqrt{\clusteringamplitude}$ can be measured. In our \LCDM{} scenario, where $D(a)$ is uniquely predicted, this translates to a factor two improvement on $\sqrt{\clusteringamplitude}$. Field-level analysis excels particularly for high $k_\mathrm{max}$ and low-noise data. Nevertheless, even if we double the noise power spectrum of our mock the improvement remains considerable by a factor $1.7$. Field-level analysis also excels in the presence of linear-order degeneracies, and we observe even stronger fractional improvements for cosmologies with a more flexible growth history (\fLCDM{} and \gLCDM{}, see subsection~\ref{sec:methods-cosmology}). There, the precision of the matter clustering amplitude improves by up to a factor three, and the parameters that describe deviations from \LCDM{} by about a factor two.

The current study considers redshift-space snapshots and does not yet contain the full information content of galaxy surveys. In particular redshift-evolution and geometric information is lacking. While $\clusteringamplitude$ and $D^2(a)$ are perfectly degenerate on a single snapshot, comparison of multiple redshifts yields information on the evolution of the growth factor. We expect field-level analysis exploits this information more effectively, because it is better able to break the degeneracy between clustering amplitude and the redshift-dependent bias. Geometric information from BAO and Alcock-Paczynski distortions is crucial to investigate the expansion history of the Universe. The measurement of the BAO scale also improves at the field-level from a better reconstruction \cite{Babic:2022dws, Babic:2024wph, Babic:2025fgv, Bayer:2026zcr}. Therefore, we expect the observed precision gains to persist as we move towards the full observational complexity. Implementation of Alcock-Paczynski distortions and a full lightcone model will allow to probe this hypothesis and quantify the gains rigorously.

To eventually realize the promises of field-level analysis on survey data some more challenges need to be addressed. It has recently been shown that the leading-order Gaussian noise term does not capture galaxy statistics sufficiently for many samples and that higher-order contributions are important. These render the likelihood intractable \cite{Rubira:2025rqo}. Further investigation into these higher-order terms and how they can be modeled efficiently is therefore required. Higher-order noise also increases the correlation length of the sampling chains, which is already considerable under the simplification of a Gaussian likelihood. A better understanding of the posterior correlation structure is crucial to optimize samplers and devise an efficient inference. And finally there are observational and systematic effects which need to be represented in the forward model.

A key science driver for stage IV cosmological surveys is the exploration of Dark Energy. Indeed, in combination with other data sets, DESI discovered hints for the departure from the most simple model of a cosmological constant \cite{DESI:2025zgx}. There is also ongoing debate on a suppression of clustering with respect to the \LCDM{} prediction from the CMB detected by weak-lensing surveys \cite{Abdalla:2022yfr, DiValentino:2020vvd}, and wether this suppression originates at early- or late-times. As our results show, field-level analysis can improve amplitude measurements from galaxy clustering, adding complementary information. Moreover, models to test structure growth and Dark Energy commonly introduces additional free parameters, and it is precisely in these extended scenarios where field-level analysis improves particularly over traditional methods by breaking degeneracies more efficiently. Moreover, it offers a self-consistent way to combine optimal BAO reconstruction with information on shape and growth. Nevertheless, the computational costs of field-level analysis will remain high for the foreseeable future, which prevents the blind exploration of all conceivable models. Guidance from theory and from traditional analyses such as power- and bispectrum will therefore be essential to best exploit the promises of field level analysis. 

\acknowledgments
I would like to thank Sveva Castello, Safak Celik, Anna Cremaschi, Francesco Conteddu, Daniel Grün, Elisabeth Krause, Noemi Anau Montel, Minh Nguyen, Ivana Nikolac, Yannik Pieper and Fabian Schmidt for enlightening discussion and Anna Cremaschi and Fabian Schmidt for their comments on the manuscript. I acknowledge funding from the Deutsche Forschungsgemeinschaft (DFG, German Research Foundation) under its Emmy-Noether Programm (STA 1955/1-1, Projektnummer 545534254). Computations were performed on the ORION cluster, maintained by the Max Planck Computing \& Data Facility. Triangle plots for field-level results have been obtained with \texttt{GetDist} \cite{Lewis:2019xzd}.

\appendix
\section{Monte Carlo Markov Chain convergence}
\label{appx:hmc}

For the field level analysis, we sample initial conditions with Hamiltonian Monte Carlo (HMC, \cite{Neal:2011mrf}) and interlace the HMC updates with slice sampling \cite{2000physics...9028N} of the cosmological parameters, noise amplitude and velocity bias coefficient. We use a diagonal mass matrix that is derived from linear-order perturbation theory. The number of steps during the HMC update is drawn from the interval $\mathcal{U}\left(2,20\right)$ and the step size from $\mathcal{U}\left(0.02, 0.12\right)$. We thin the chains by a factor ten in parameters and by a factor of 100 in $\hat{s}$ to save storage and provide easier file handling.

For each scenario, we initialize eight chains from different random seeds. Four start from the ground truth. For the other four, we draw random starting points for the cosmological parameters from a flat interval that covers the $\pm 3\sigma$ region of the Fisher forecast. For this second set, we also draw random starting points for $\hat{s}$ from the prior and divide their amplitude by ten. Drawing from an under-dispersed priors helps to prevent the chains from getting stuck in local extrema during the warm-up phase. Sampling of the randomly initialized chains proceeds in two steps; during warm-up we keep cosmology parameters fixed and only sample the phases. To aid convergence, we perform annealing during the warm-up phase and fix the noise amplitude $\noiseamplitude$ to a high value (initially 2.0 for the fiducial mock and 2.4 for the noisy mock) which we reduce in steps of 0.2 every 1.000 samples. Once the annealing has finished, we switch to joint sampling of phases and cosmological parameters. The acceptance rate in this phase is around $0.5$.

We remove a burn-in period from each chain upon visual inspection, then compute the effective sample size per chain and the Gelman-Rubin criterion $R-1$ over the ensemble. Our stopping criteria are 100 effective samples in the cosmological parameters and noise amplitude accumulated over all chains and $R-1 \leq 0.05$. We find that the noise amplitude $\noiseamplitude$ and the clustering amplitude parametrized by $\alpha$ exhibit the longest correlation length, while in comparison the initial conditions decorrelate more quickly. The effective sample size and Gelman-Rubin criterion of all analyses presented in this paper are summarized in table~\ref{tab:chain-convergence}.
  
\begin{table}
	\centering
\begin{tabular} {c | c | c | c | c | c}
  & \multicolumn{4}{c|}{fiducial mock} & \multicolumn{1}{c}{noisy mock} \\ 
 \hline
 & \LCDM{} &  \makecell{\LCDM{},\\ 3rd order bias} & \gLCDM{} & \fLCDM{} & \LCDM{}\\ 
 \hline
 \hline
$\alpha$  &  \makecell{170 \\ 0.015} & \makecell{320 \\ 0.034} & \makecell{190 \\ 0.018} & \makecell{170 \\ 0.022} & \makecell{120 \\ 0.038}\\
\hline
$\omega_\matter$ &  \makecell{1.300 \\ 0.003} & \makecell{2.200 \\ 0.005} & \makecell{1.300 \\ 0.001} & \makecell{1.600\\ 0.001}& \makecell{1.000 \\ 0.007}\\
\hline
$\gamma$  &-- & -- & \makecell{480 \\ 0.009 } & --  &  -- \\
\hline
$f$  &-- & -- & -- & \makecell{440\\ 0.012} & -- \\
\hline
$\noiseamplitude$ &  \makecell{100 \\ 0.026} & \makecell{86 \\ 0.078}& \makecell{140 \\ 0.044} & \makecell{96 \\ 0.039}& \makecell{120 \\ 0.041} \\
\hline
$\beta_{\partial_{||}\sigma}$ &  \makecell{1.900 \\ 0.002} & \makecell{2.600 \\ 0.007} & \makecell{2.100 \\ 0.002} & \makecell{1.800\\ 0.003} & \makecell{1.100 \\ 0.006}  \\
\end{tabular}
\caption{Effective sample size (upper set of numbers) and Gelman-Rubin criterion $R-1$ (lower set of numbers) for all field-level analyses presented. Bias coefficients $b_\op$ are marginalized analytically and hence do not appear here.}
\label{tab:chain-convergence}
\end{table}

\begin{figure}
\centering
\includegraphics[]{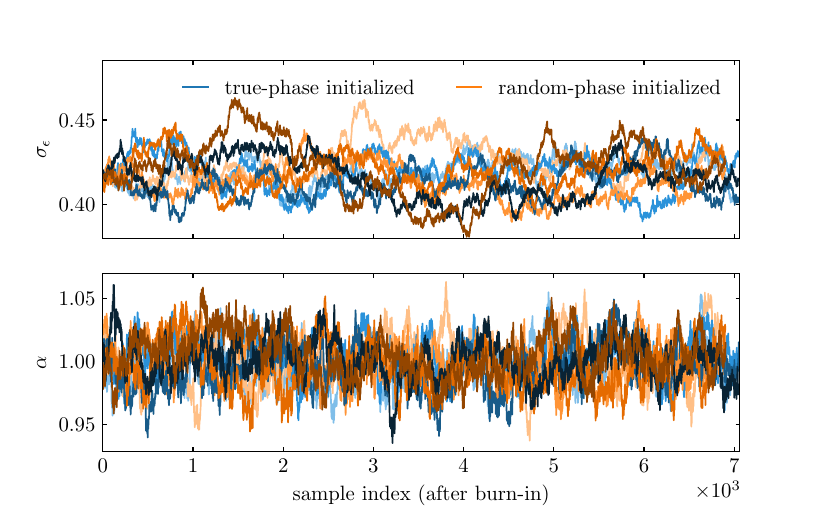}
\caption{Trace plot of all eight sampling chains from the \LCDM{}, 3rd order bias scenario on the fiducial mock. Of all analyses performed, this one has the longest correlation lengths in particular in the parameters shown here, $\alpha \propto \sqrt{\clusteringamplitude}$ and $\noiseamplitude$. Chains that start from the ground truth are shown in blue shades, those that start from a random initialization of cosmological parameters and initial conditions in orange. The burn-in phases is removed and the chains are thinned by a factor 10.}
\label{fig:appx-convergence-trace}
\end{figure}

The \LCDM{} analysis with third order bias on the fiducial mock exhibits particularly long correlations that challenge our convergence targets for $\noiseamplitude$, see table~\ref{tab:chain-convergence}. This is even though the chains collected nearly twice as many samples compared to the other scenarios. The trace plots of $\noiseamplitude$ and the clustering amplitude $\alpha$ are shown in figure~\ref{fig:appx-convergence-trace}, and we further investigate these chains for pathologies such as multi-modality. Excluding chains one-by-one, we find that $R-1$ of $\noiseamplitude$ is driven by one particular chain (in figure~\ref{fig:appx-convergence-trace} the chain in the palest orange shade), and there is no comparable effect on the other parameters. Removing this chain leaves the parameter standard deviations essentially unchanged. The likelihood values along all chains are fully consistent among each other whereas differences would indicate isolated local extrema. We repeat the phase plot in figure~\ref{fig:results-violin} to compare the samples initial conditions of each chain against the whole ensemble and find essentially no difference. In summary, we find no signs of pathologies, and we expect that running these chains longer, they would reach the convergence criteria also in $\noiseamplitude$. Still, in all other parameters the targets are already achieved, and we expect only a very minor impact on the final parameter contours. Further optimization of the HMC parameters, in particular the number and size of integration steps, could help to bring down the correlation length but would require start all chains from scratch. Clearly, this example particularly highlights the challenges associated with numerical sampling the field-level posterior and the need for further optimization of the sampler to achieve more efficient work flows.

\section{Fisher forecast numerical implementation}
\label{appx:fisher}

Covariances and derivatives for the Fisher forecast are estimated from an ensemble of \LEFTfield{} simulations. For the covariance, we generate $4\times 10^5$ realizations with random initial conditions and noise and measure the power spectrum multipoles $l=0,2,4$ and the bispectrum monopole on each sample. This data vector has 1265 entries, corresponding to a Hartlap factor \cite{Kaufman1967, Hartlap:2006kj} of $h = 1.003$. We do not apply the Hartlap factor since it would only increase the contours. We estimate the covariance, including off-diagonal elements from this ensemble. To verify the number of samples used in the covariance estimation, we divide the ensemble into four random subsets of half the size and repeat the Fisher forecast with the covariances computed from the subset. We find that all one-dimensional parameter intervals are stable to less than $1\,\%$.

\begin{table}
\centering
\begin{tabular}{c | c | c | c | c | c }
	& & \multicolumn{2}{c|}{sample size} & \multicolumn{2}{c}{step size} \\
scenario & \# simulations &  cosmology & bias & cosmology & bias \\
\hline
\LCDM{}, 2nd order bias & 7430 & $\leq 0.4\,\%$ & $\leq 0.4\,\%$ & $\leq 0.2\,\%$ & $\leq 0.2\,\%$ \\
\LCDM{}, 3rd order bias & 6410 & $\leq 1.0 \,\%$ & $\leq 3.5\,\%$ & $\leq 0.6\,\%$ & $\leq 2.5\,\%$ \\
\gLCDM{}, 2nd order bias & 12600 &  $\leq 0.4\,\%$ & $\leq 1.0\,\%$ &  $\leq 0.2\,\%$ & $\leq 0.4\,\%$ \\
\fLCDM{}, 2nd order bias & 12200 & $\leq 0.4\,\%$ & $\leq 1.0\,\%$  & $\leq 0.7\,\%$ & $\leq 1.4\,\%$
\end{tabular}
\caption{Estimated numerical uncertainty in the Fisher forecast from averaging derivatives over a finite set of initial conditions and noise realizations, and from the finite step size. We only list analyses on the fiducial mock, and we rescale derivatives for the noisy mock. Taken together with numerical errors from the covariance, we estimate the numerical accuracy of one-dimensional Fisher intervals on cosmological parameters to be better than $2\,\%$ ($3\,\%$ for \LCDM{} with 3rd order bias).}
\label{tab:fisher-numerical-accuracy}
\end{table}

We estimate derivatives from finite differences by running \LEFTfield{} with a small step in one parameter at a time, and we average the estimate over an ensemble of initial conditions and noise realizations. We estimate the impact of the finite sample size by comparing our result to two randomly selected sub-samples that contain only half the simulations. The step size $\epsilon$ for computing the derivatives is set to roughly match the $1\sigma$ interval in each parameter. We verify this  choice by comparing to two sets of derivatives computed with a smaller and larger step size, $\epsilon' = \epsilon \pm 0.5\epsilon$. The comparison between original an smaller/larger stepsizes is performed only for a subset of $1.000$. By starting these simulations from the same random seed for the initial conditions and noise draws, we can nevertheless disentangle the effect of derivative stepsize from cosmic variance. We report the results of these tests in table~\ref{tab:fisher-numerical-accuracy}. While we compute separate covariances for the fiducial and the noisy mock, we save computational costs by rescaling the derivatives in $\sigma$ for the noisy mock.

Combining the numerical uncertainty from the finite sample size to estimate covariances and derivatives and from the finite step size to estimate derivatives, we find that one-dimensional contours on the cosmological parameters are impacted by less than $2\,\%$ with the exception of the 3rd order bias \LCDM{} analysis on the fiducial mock, where the accuracy is $\lesssim 3\,\%$. 

\section{Bias marginalization and conditional bias posterior}
\label{appx:parsampling}

Bias coefficients enter the field-level likelihood (eq.~\ref{eq:intro-likelihood}) quadratically, and it is possible to marginalize over them analytically \cite{Elsner:2019rql}. To this end, we split the bias operators in two non-overlapping sets, $\{\op\}_\explicit$ and $\{\op\}_\marginalized$; the former will be sampled explicitly the latter marginalized analytically. With this split, we decomposed the deterministic part of the bias as
\begin{equation}
	\tilde{\delta}_\tracerdet(\vec{k}) = \tilde{\mu}(\vec{k}) + \sum_{{\op\in\{\op\}_\marginalized}} b_\op\, \tilde{\op}(\vec{k})\,, 
	\quad \mathrm{where}\quad
	\tilde{\mu}\left(\vec{k}\right) = \sum_{\op\in\{\op\}_\explicit}\,b_\op\, \tilde{\op}(\vec{k}) .
	\label{eq:parsampling-split}
\end{equation}
We can now write the posterior as
\begin{align}
-\ln \prob\left(\left. \hat{s}, \theta, b_\op \right| \hat{\delta}_\tracer\right) 
= &
\sum_{\vec{k}\neq 0}^{k_\mathrm{max}} \ln \noiseamplitude -\frac{C}{2}
- \sum_{\{\op\}_\marginalized} b_\op\,  B_\op  
+ \sum_{\{\op\}_\marginalized, \{\op'\}_\marginalized} \frac{b_\op b_{\op'}}{2} A_{\op\op'}
 \nonumber  \\[.5\baselineskip]
&-\ln \prob\left(\hat{s} | \theta \right) - \ln \prob\left(\theta , \{b_\op\}_\explicit\right) - \ln \prob\left(\{b_\op\}_\marginalized\right)
\label{eq:parsampling-posterior}
\end{align}
where we use the following shorthand notations,
\begin{align}
	C \left(\hat{s},\theta,\{b_\op\}_\explicit\right)  &= \sum_{\vec{k}\neq 0}^{k_\mathrm{max}} \frac{\left|\tilde{\delta}_\tracer(\vec{k}) - \tilde{\mu}(\vec{k}) \right|^2}{\noiseamplitude^2}\,,
	\\
	B_\op \left(\hat{s},\theta,\{b_\op\}_\explicit\right)  &= \sum_{\vec{k}\neq 0}^{k_\mathrm{max}}  \frac{\Re\left\{\left[\tilde{\delta}_\tracer(\vec{k}) - \tilde{\mu}(\vec{k})\right]\op^{*}\right\}}{\noiseamplitude^2}\,,
	\label{eq:parsampling-B}
	\\
	A_{\op \op'} \left(\hat{s},\theta,\{b_\op\}_\explicit\right)  &= \sum_{\vec{k}\neq 0}^{k_\mathrm{max}} \frac{\op(\vec{k})\, \op'^*(\vec{k})}{\noiseamplitude} + \delta_{\op,\op'} \left(C^{-1}_\mathrm{prior}\right)_{\op\op'}\,.
	\label{eq:parsampling-A}
\end{align}
That is, $C$ is a scalar, $B$ a vector whose dimensionality corresponds to the number of marginalized operators, and $A$ is a square matrix. We assume that the prior factorizes between explicit and marginalized parameters, and that the latter is given by a normal distribution with diagonal covariance $C_\mathrm{prior}$, see also table~\ref{tab:parameters}. Analytic marginalization over $\{b_\op\}_\marginalized$ yields \cite{Elsner:2019rql},
\begin{align}
	-\ln \prob\left( \left. \hat{s}, \theta, \{b_\op\}_\explicit \right| \tilde{\delta}_\tracerdet \right) 
	 &= -\ln \left[ \left(\prod_{\op\in\{\op\}_\marginalized} \int d b_\op \right) \prob\left( \left. \hat{s}, \theta, b_\op \right| \tilde{\delta}_\tracerdet \right)  \right]
	\nonumber \\
	&=\frac{1}{2} \left\{
	C + \sum_{\vec{k}\neq 0}^{k_\mathrm{max}} \ln \noiseamplitude^2
	+ \ln \left|A_{\op\op'}\right|
	- \sum_{{\op,\op'\in\{\op\}_\marginalized}} B_\op\, \left[A^{-1}\right]_{\op \op'}\, B_{\op'}
	\right\} \nonumber \\[.5\baselineskip]
	&-\ln \prob\left(\hat{s} | \theta \right) - \ln \prob\left(\theta , \{b_\op\}_\explicit\right) - \ln \prob\left(\{b_\op\}_\marginalized\right) + \mathrm{const.}\,.
\label{eq:parsampling-margPost}
\end{align}

Analytic bias marginalization has been shown to reduce the correlation length in the Markov chains and lead to more efficient sampling \cite{Kostic:2023arx}. Hence, for the analyses presented here, we marginalize over all bias coefficients analytically. Yet, to better understand the prominent posterior degeneracies and gain intuition how field-level analysis differs from the power- and bispectrum, access to the full posterior including all bias coefficients is desirable. We can write the full posterior as
\begin{equation}
\prob\left(\left. \hat{s}, \theta, \{b_\op\}_\marginalized, \{b_\op\}_\explicit \right| \hat{\delta}_\tracer \right)
= \prob\left(\left. \{b_\op\}_\marginalized, \right| \hat{s}, \theta, \{b_\op\}_\explicit, \hat{\delta}_\tracer, \right) \,
\prob\left(\left. \hat{s}, \theta, \{b_\op\}_\explicit \right| \hat{\delta}_\tracer \right) \,,
\end{equation}
where the last term on the right-hand side is the marginalized posterior from eq.~(\ref{eq:parsampling-margPost}). This means, once we have obtained samples from the marginalized posterior, we can draw from the conditional distribution for the marginalized parameters to obtain samples from the full posterior. The conditional posterior for the marginalized bias coefficients is obtained by evaluating eq.~(\ref{eq:parsampling-posterior}) for the given values of $\hat{s}$, $\theta$ and $\{b_\op\}_\explicit$. Only the last two terms in the first line of eq.~(\ref{eq:parsampling-posterior}) and the final term of the second line depend on $\{b_\op\}_\marginalized$. Completing the square yields then yields the distribution of the marginalized parameters,
\begin{equation}
 \prob\left(\left. \{b_\op\}_\marginalized, \right| \hat{s}, \theta, \{b_\op\}_\explicit, \hat{\delta}_\tracer, \right) = \mathcal{N}\left( \left. \left(B\, A^{-1}\right)^T \right| A + C_\mathrm{prior}^{-1} \right) \,,
\end{equation}
where $\mathcal{N}$ denotes a multivariate normal distribution and $A$, $B$ and $C$ are evaluated for the current values of $\hat{s}$, $\theta$ and $\{b_\op\}_\explicit$.

\bibliographystyle{JHEP} 
\bibliography{fli-redshiftspace.bib}
\end{document}